\documentclass[journal,10pt]{IEEEtran}

\usepackage{oldlfont}
\usepackage{psfrag}
\usepackage{latexsym}
\usepackage{graphics}
\usepackage{epsfig}
\usepackage{amsmath}
\usepackage{amssymb}
\usepackage{array}
\usepackage{color}

\begin{document}

\title{Synchronization recovery and state model reduction for 
soft decoding of variable length~codes}
\author{Simon Malinowski, Herv\'e J\'egou, and Christine Guillemot}

\maketitle

\newcommand{\equaldef}
{=^{\hspace{-0.25cm} \Delta}}

\def\A{{\mathcal A}}
\def\B{{\mathcal B}}
\def\N{{\mathcal N}}
\def\Cset{{\mathcal C}}
\def\Pset{{\mathcal P}}
\def\D{{\mathcal D}}
\def\T{{\mathcal T}}
\def\Z{{\mathcal Z}}
\def\I{{\mathcal I}}
\def\R{{r}}
\def\Ri{{\mathcal R}_i}
\def\S{{\mathbf S}}
\def\s{{\mathbf s}}
\def\X{{\mathbf X}}
\def\x{{\mathbf x}}
\def\Y{{\mathbf Y}}
\def\e{{\mathbf e}}
\def\P{{\mathbb P}}
\def\b{\overline{b}}
\def\l{\overline{l}}
\def\voidsequence{\varepsilon}

\newcounter{example}
\setcounter{example}{0}

\newcounter{definition}
\setcounter{definition}{0}

\newcounter{property}
\setcounter{property}{0}

\newcounter{theorem}
\setcounter{theorem}{0}

\def\theexample{\arabic{example}}
\def\thedefinition{\arabic{definition}}
\def\theproperty{\arabic{property}}

\newenvironment{example}%
{\refstepcounter{example} 
\smallskip
\noindent {\it Example \theexample:} \bgroup \sl}%
{\egroup \smallskip}

\newenvironment{definition}%
{\refstepcounter{definition} 
\smallskip
\noindent {\bf Definition \thedefinition:} \bgroup}%
{\egroup \smallskip}

\newenvironment{property}%
{\refstepcounter{property} 
\smallskip
\noindent {\it Property \theproperty:} \bgroup \sl}%
{\egroup \smallskip}

\newenvironment{theorem}%
{\refstepcounter{theorem} 
\smallskip
\noindent {\it Theorem \thetheorem:} \bgroup \sl}%
{\egroup}

 \newcommand{\col}[1]
   {\colorbox{yellow}{\textcolor{red}{#1}}}


\begin{abstract} 

Variable length codes (VLCs) exhibit de-synchronization problems when transmitted over noisy channels. Trellis decoding techniques 
based on Maximum A Posteriori (MAP) estimators are often used to minimize the error rate on the estimated sequence.  If the number 
of symbols and/or bits transmitted are known by the decoder, termination constraints can be incorporated in the decoding process. 
All the paths in the trellis which do not lead to a valid sequence length are suppressed. This paper presents an analytic method 
to assess the expected error resilience of a VLC when trellis decoding with a sequence length constraint is used.  The approach 
is based on the computation, for a given code, of the amount of information brought by the constraint. It is then shown that this 
quantity as well as the probability that the VLC decoder does not re-synchronize in a strict sense, are not significantly altered 
by appropriate trellis states aggregation. This proves that the performance obtained by running a length-constrained Viterbi decoder on 
aggregated state models approaches the one obtained with the bit/symbol trellis, with a significantly reduced complexity. 
It is then shown that the complexity can be further decreased 
by projecting the state model on two state models of reduced size.
\end{abstract}

\section{Introduction} 
VLCs are widely used in compression systems due to their high compression efficiency. One drawback of VLCs is their high sensitivity 
to errors. A single bit error may lead to the de-synchronization of the decoder.  Nevertheless, many VLCs exhibit self-synchronization 
properties. The authors in~\cite{FeR84} show such properties for some binary Huffman codes.  The error recovery 
properties of VLCs have also been studied in~\cite{MaR85}, where a method to compute the so-called \emph{expected error span} $E_s$ 
(i.e. the expected number of source symbols on which a single bit error propagates), has been proposed.  The same quantity has been 
called {\it mean error propagation length} (MEPL) in~\cite{ZhZ02}.  The authors in~\cite{MoL87} consider the variance of the error 
propagation length (VEPL) to assess the resilience of a code with hard decoding techniques.  In~\cite{SwD95}, the method of~\cite{MaR85} 
is extended to compute the so-called \emph{synchronization gain/loss}, i.e. the probability that the number of symbols in the 
transmitted and decoded sequences differ by a given amount $\Delta S$ when a single bit error occurs during the transmission.  
Note that various VLC constructions have also been proposed to improve the self-synchronization 
properties of the codes~\cite{MoA86},~\cite{Tit96},~\cite{FJT03}.  The author in~\cite{Tit96} introduces a method 
to construct prefix-free self-synchronizing VLCs called T-codes. The synchronization property of these codes is analyzed 
in~\cite{Tit97} in terms of the {\it expected synchronization delay}.

VLC soft decoding techniques based on MAP (or MMSE) estimators have also been considered to minimize the error rates (or distortion) 
observed on the decoded sequences.  The approaches essentially differ in the optimization metrics as well as in the assumptions 
made on the source model and on the information available at the decoder.  These assumptions lead to different trellis structures 
on which the estimation or soft-decoding algorithms are run. Two main types of trellises are considered to estimate the sequence 
of emitted symbols from the received noisy bitstream: the bit-level trellis proposed in~\cite{Bal97} and the bit/symbol trellis. 
The bit-level trellis leads to low decoding complexity. However, it does not allow the exploitation of extra information, such as 
the number of emitted symbols. It hence suffers from some suboptimality. If the knowledge 
of the number of emitted symbols is available at the decoder, the problem is referred to as \emph{soft decoding with length constraint}  
and is addressed, e.g., in~\cite{MuF99}\cite{MiP98b}\cite{gfgr00}\cite{KlT05}. This problem has led to the introduction of the bit/symbol 
trellis in~\cite{BaH00e}. This trellis can optimally exploit such constraints, leading to optimal performance in terms of error resilience. 
Nevertheless, the number of states of the bit/symbol trellis is a quadratic function of the sequence length. The corresponding complexity 
is actually not tractable for typical sequence lengths. In order to overcome this complexity hurdle, most authors apply suboptimal estimation 
methods on this optimal state model such as sequential decoding~\cite{KlT05}\cite{MuF98}\cite{BKK01}.

This paper presents a method to assess the error resilience of VLCs when trellis decoding with length constraint is used at the decoder side.   
The approach is based on the concept of gain polynomials defined on error state diagrams introduced in \cite{MaR85} and \cite{SwD95}.  
The method introduced in~\cite{SwD95} to compute the synchronization gain/loss is first recalled.  This method is then extended to the 
case of a symbol sequence of length $L(S)$ being sent over a binary symmetrical channel (BSC) of a given crossover probability. 
The derivation is inspired from the matricial method described in~\cite{ZhZ02}. It has been shown in \cite{Wei03}\cite{ThK03} that 
the Markovian property of a source can be easily integrated in the source model by expanding the state model by a constant factor. 
We thus restrict the analysis to memoryless sources.  It is shown that for VLCs, the probability mass function (p.m.f.) of the synchronization 
gain/loss is a key indicator of the error resilience of such codes when soft decoding with length constraint is 
applied at the decoder side.  The p.m.f. of the gain/loss allows the computation of the probability that the symbol length of the decoded 
sequence is equal to $L(S)$, i.e. the probability that the decoder resynchronizes in the strict-sense (no gain nor loss of symbols during 
the transmission). This quantity is given by $\P(\Delta S = 0)$.  The length constraint is used to discard all decoded sequences which do 
not satisfy the constraint $\Delta S = 0$. If $P(\Delta S = 0)$ is high, the number of ``de-synchronized'' sequences which will be 
discarded will be high. This results in increasing the likelihood of the correct sequence, hence in decreasing the decoding error rate.  
The entropy of the p.m.f. of the gain/loss represents the amount of information that the length constraint brings to the decoder. These 
two quantities ($\P(\Delta S = 0)$ and $H(\Delta S)$) are shown to better predict the relative decoding performance of VLCs when soft decoding with a 
length constraint is used,than the MEPL 
and VEPL measures (these measures are appropriate when hard decoding is used). 
Note that, in the following, the term MEPL will be used to refer to the expectation of the error propagation length. 

This analysis is then used in Section~\ref{sec:statemodel} to assess the performance of MAP decoding on the aggregated state models 
proposed in \cite{JMG05}, for jointly typical source/channel realizations. The aggregated state model is defined by both the internal 
state of the VLC decoder (i.e., the internal node of the VLC codetree) and the remainder of the Euclidean  division of the symbol clock 
values by a fixed parameter called $T$. This model aggregates states of the bit/symbol trellis which differs by multiple of $T$symbol clock instants. 
The parameter $T$ controls the trade-off between estimation accuracy and decoding complexity. The choice of this parameter has 
indeed an impact on the quantity of information brought by the length constraint on the corresponding trellis. It is shown that the 
probability that the VLC decoder does not re-synchronize in a strict sense, as well as the entropy of the constraint, are not significantly 
altered by aggregating states, provided that the aggregation parameter $T$ is greater than or equal to a threshold.  An upper bound of this 
threshold is derived according to the analysis of Section~\ref{sec:synchro}.  This proves that the performance obtained by running a 
length-constrained Viterbi decoder on the aggregated trellis closely approaches the performance obtained on the bit/symbol trellis, 
with a significantly reduced complexity.  Finally, it is shown in Section~\ref{sec:mtd} that the decoding complexity can be further 
reduced by considering separate estimations on trellises of smaller dimensions, whose parameters $T_1$ and $T_2$ are relatively prime.  
If the two sequence estimates are not equal, the decoding on a trellis of parameter $T_1 \times T_2$ is then computed.  The equivalence in terms
of decoding performance between this approach, referred to as combined trellis decoding, and the decoding on a trellis of parameter~$T_1 \times T_2$ is proved for 
the MAP criterion, i.e. for the Viterbi algorithm~\cite{Vit67}.  

\section{Link between VLC synchronization recovery properties and soft decoding performance with a length constraint}
\label{sec:synchro}

Let $\S=S_1, ... S_t, ...S_{L(\S)}$ be a sequence of $L(\S)$ symbols. This sequence is encoded with a VLC $\Cset$,  
producing a bitstream $\X=X_1,... X_k, ... X_{L(\X)}$ of length $L(\X)$.  This bitstream is modulated using a binary phase 
shift keying (BPSK) modulation and is transmitted over an additive white Gaussian noise (AWGN) channel, without any channel 
protection. The channel is characterized by its signal to noise ratio, denoted ${E_b}/{N_0}$ and expressed in decibels (dB).   
Note that we reserve block capital letters to represent random variables and small letters to represent their corresponding realizations.  
In this paper, the term \emph{polynomial} refers to expressions of the form $\sum_{i \in {\mathbb Z}} a_i x^i$, where $x$ denotes 
the variable and $a_i$ are polynomial coefficients. Hence, we include in this terminology either polynomial series (with an infinite 
number of non null coefficients) or finite length polynomials (such as $\exists N \in {\mathbb N} \, \, | \, \, \forall n>N,\ a_{-i}=a_i = 0$), both with negative powers. 

\subsection{The \emph{gain/loss} behavior of a variable length code}

A method to compute the so-called \emph{expected error span} $E_s$ following a single bit error has been introduced
in~\cite{MaR85}. This method relies on an error state diagram which represents the states of the decoder when the encoder 
is in the root node. Hence, the error state diagram includes the internal states of the decoder, i.e. the internal nodes 
of the VLC, plus two states which represent the \emph{loss of synchronization state} $n_l$ and the \emph{return to synchronization state} 
$n_s$ respectively. Therefore, the set of states of the diagram is $\{n_l, n_{\alpha_1}, n_{\alpha_2}, ... ,n_s \}$, where 
the set $\{\alpha_1, \alpha_2, ...\}$ represents the set of prefixes of the VLC. The state $n_s$ of the error state diagram 
corresponds to a return of both encoder and decoder automata to the root node of the code tree.  However, this state may not 
correspond to a \emph{strict sense} synchronization. In other words, the number of decoded symbols may be different from the 
number of emitted ones. The branches of the error state diagram represent the transitions between two states of the decoder 
when a single source symbol has been emitted by the encoder.  They are labeled by an indeterminate variable $z$ which corresponds 
to the encoding of one source symbol.  Hence, the gain along each edge is the probability of the transition associated with that edge 
multiplied by $z$.  In that case, the gain on the diagram from $n_l$ to $n_s$ (i.e. the transfer function between $n_l$ and $n_s$) is a 
polynomial of the variable $z$ such that the coefficient of $z^i$ is the probability that the considered VLC resynchronizes after 
exactly $i$ source symbols following the bit error. Evaluating the derivative of the gain polynomial at $1$ provides the expected error span $E_s$. 

The branch labeling of the error state diagram has been extended in~\cite{SwD95} so that the gain polynomial informs about the 
difference, caused by a single bit error, between the number of emitted and decoded symbols, after hard decoding of the received 
bitstream. This quantity, denoted $\Delta S$, is referred to as the \emph{gain/loss}. In order to evaluate the p.m.f. of the random 
variable $\Delta S$, a new variable $y^j$ is introduced in the branch labeling of the error state diagram. The exponent $j$ 
represents the number of extra output symbols for each input symbol. Hence, the corresponding gain polynomial $G(y,z)$ is function 
of both variables $y$ and $z$.  Evaluating this polynomial at $z=1$ gives a polynomial in $y$ only. For sake of 
clarity, we simply denote this polynomial as 
\begin{equation}
G(y) = G(y,z)|_{z=1}. 
\end{equation}
The coefficient of $y^i$ in the polynomial $G(y)$ gives the probability $\P(\Delta S = i)$ following one bit error.
Note that $i$ can be negative if the decoded sequence is longer than the encoded one. In this section, we focus on the behavior 
of the polynomial $G(y)$.  Since the variable $z$ is not necessary, we compute directly the state diagram for $z=1$. 

Let $\overline{H}$ be the transition matrix corresponding to the error state diagram.
\begin{equation}
\overline{H} = \left (
       \begin{array}{cccc}
          \P(n_l | n_l) & \P(n_{\alpha_1}|n_l) & \cdots & \P(n_s|n_l) \\
          \P(n_l|n_{\alpha_1}) & \P(n_{\alpha_1}|n_{\alpha_1}) & \cdots & \P(n_s|n_{\alpha_1}) \\
          \vdots & \vdots & \ddots & \vdots \\
          \P(n_l|n_s) & \P(n_{\alpha_1}|n_s) & \cdots & \P(n_s|n_s) \\
   \end{array}
   \right )
\end{equation}
where $\P(n_{\alpha_i}|n_{\alpha_j})$ represents the probability to go to state $n_{\alpha_i}$ from state $n_{\alpha_j}$.
Let us call $h_{i,j}^k(y)$ the element at row $i$ and column $j$ of the matrix $\overline{H}^k$. Note that $h_{i,j}^k(1)$ 
is the probability to go from state $n_{\alpha_i}$ to state $n_{\alpha_j}$ in $k$ stages, i.e. after the encoding of $k$ 
source symbols. The top right elements of the matrices $\overline{H}$ and $\overline{H}^k$ are respectively denoted $h(y)=\P(n_s|n_l)$ and $h^k(y)$. 
The gain polynomial $G(y)$ can then be written as 
\begin{equation}
\label{equ:gain_polynomial}
G(y) = \sum_{ k \in \mathbb{N}^*} h^k(y). 
\end{equation}
Hence, the gain polynomial $G(y)$ is obtained as the top right element of the matrix $(\overline{I}-\overline{H})^{-1}$, 
where $\overline{I}$ denotes the identity matrix of the same dimensions as $\overline{H}$. 
Note that this property holds if $(\overline{I}-\overline{H})^{-1}$ exists.

Let us consider the $5$-symbol source and the 16 VLCs used in~\cite{ZhZ02} to illustrate these concepts.  
The probability of this source as well as the different codes are reproduced in Table~\ref{tab:codesdef}.
These codes have the same mean description length of $2.2$ bits per symbols.  

\begin{table}
\caption{Source and Codes from~\cite{ZhZ02} used in this paper.}
\label{tab:codesdef}
$$
\begin{array}{|c|ccccc|}
\hline
a_i        &  a_1 & a_2 & a_3 & a_4 & a_5  \\
\hline
\P(a_i)  = &  0.4 & 0.2 & 0.2 & 0.1  & 0.1 \\
\hline
\Cset_{1}  &  00 & 01  & 10  & 110  & 111  \\
\Cset_{2}  &  00 & 01  & 11  & 100  & 101  \\
\Cset_{3}  &  00 & 10  & 11  & 010  & 011  \\
\Cset_{4}  &  01 & 00  & 10  & 110  & 111  \\
\Cset_{5}  &  01 & 00  & 11  & 100  & 101  \\ 
\Cset_{6}  &  01 & 10  & 11  & 000  & 001  \\
\Cset_{7}  &  0  & 10  & 110 & 1110 & 1111 \\
\Cset_{8}  &  0  & 10  & 111 & 1100 & 1101 \\
\Cset_{9}  &  0  & 11  & 100 & 1010 & 1011 \\
\Cset_{10} &  0  & 11  & 101 & 1000 & 1001 \\
\Cset_{11} &  0  & 100 & 101 & 110  & 111  \\
\Cset_{12} &  0  & 100 & 110 & 101  & 111  \\
\Cset_{13} &  0  & 100 & 111 & 110  & 101  \\
\Cset_{14} &  0  & 101 & 110 & 100  & 111  \\
\Cset_{15} &  0  & 101 & 111 & 100  & 110  \\
\Cset_{16} &  0  & 110 & 111 & 100  & 101  \\
\hline
\end{array}
$$
\end{table}

\begin{figure}[t]
\begin{center}
    \includegraphics[width=8cm]{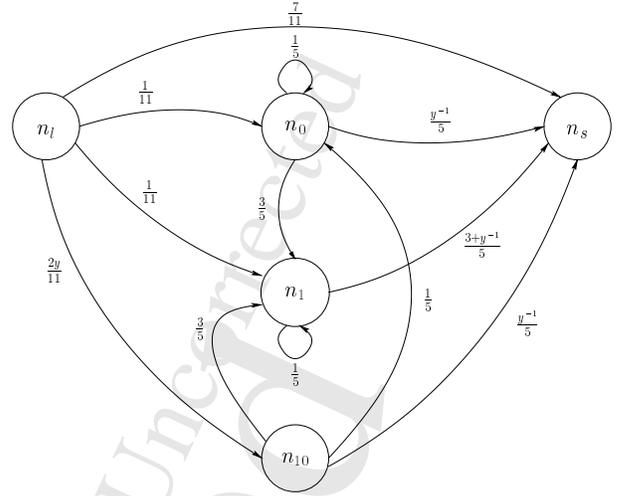}
\end{center}
\caption{Error state diagram of \cite{SwD95} for the code ${\mathcal C}_{5}$. 
The transition probabilities are denoted next to the branches.}
\label{fig:esd_C5}
\end{figure}

\begin{example}
Let us consider the code $\Cset_{5}$. Its state diagram is depicted in Fig.~\ref{fig:esd_C5}.  
The transition matrix derived from the previous guidelines, i.e. by setting the variable z to 1 in the extended diagram of \cite{SwD95}, is given by 
$$ \overline{H}_{\Cset_{5}} = \left (
           \begin{array}{ccccc}
           0 & \frac{1}{11} & \frac{1}{11} & \frac{2y}{11} & \frac{7}{11} \\
           0 & \frac{1}{5} & \frac{3}{5} & 0 & \frac{y^{-1}}{5} \\
           0 & 0  & \frac{1}{5} & 0 & \frac{3+y^{-1}}{5} \\
           0 & \frac{1}{5} & \frac{3}{5} & 0 & \frac{y^{-1}}{5} \\
           0 & 0 & 0 & 0 & 0 \\
           \end{array}
           \right ).$$
This leads to $G(y)= 0.0625y^{-1} + 0.8352 + 0.1023y$, which also means that
\begin{equation}
\left\{
\begin{array}{ll}
\P(\Delta S = -1)  & = 0.1023 \\
\P(\Delta S = 0)  & = 0.8352 \\
\P(\Delta S = 1) & = 0.0625. 
\end{array}
\right.
\end{equation}
\end{example}

\subsection{Extension to the BSC}
\label{sec:deltaSbsc}

Let us recall that $\Delta S$ corresponds to the gain/loss engendered 
by a \emph{single} bit error.
We propose here to estimate $\P(\Delta S = i)$ for a sequence of $L(\S)$ symbols that has been
sent through a BSC of crossover probability $p$ (equals to the bit error rate).
Since in this section the VLC decoder is assumed to be a classical hard decoder, 
the analysis is also valid on an AWGN channel characterized by its signal to noise ratio $E_b/N_0$
by taking $\textrm{p} = \frac{1}{2} \textrm{erfc}\left(\sqrt{\frac{E_b}{N_0}}\right)$. 

For a sequence of $L(\S)$ symbols, the bitstream length $L(\X)$ lies in the interval of integers 
$\I = \{L(\S) \times l_m,\dots,L(\S) \times \, l_M\}$, 
where $l_m$ and $l_M$ respectively denote the lengths of the shortest and longest codewords.
Let $E$ denote the random variable corresponding to the number of errors after the hard decoding
of the received bitstream $\Y$.
For $i \in {\mathbb Z}$, the probability $\P(\Delta S = i)$ is given by
\begin{equation}
\P(\Delta S = i)=\sum_{e \in {\mathbb N}} \P(\Delta S = i \, | \, E=e)\P(E=e).
\label{equ:DeltaS}
\end{equation}
For $e \in {\mathbb N}$, the probability $\P(E=e)$ can be expressed as 
\begin{align}
\P(E=e) = \sum_{k \in \I} \P(E=e|&L(\X)=k)\P(L(\X)=k) \\ & \textrm{if } e \leq L(S) \times l_M \\
\textrm{otherwise }\P(E=e) = 0, & 
\label{equ:PEe}
\end{align}
where the quantities $\P(E=e|L(\X) = k)$ only depend on the signal to noise ratio 
and are equal to 
\begin{equation}
\left\{
\begin{array}{lll}
\P(E=e|L(\X) = k) & = {k \choose e} \, \textrm{p}^e \, (1-\textrm{p})^{i-e} & \textrm{if } e \leq k \\
\P(E=e|L(\X) = k) & = 0 & \textrm{if } e > k.
\end{array}
\right. 
\label{equ:pel}
\end{equation}
For every $k \in \I$, the probability $\P(L(\X) = k)$ is calculated from the source statistics and the code ${\mathcal C}$ structure.

To calculate $\P(\Delta S = i)$ according to Eqn.~\ref{equ:DeltaS}, we now need to compute
the quantities $\P(\Delta S = i \, | \, E=e)$.
For that purpose, let us now assume that the decoder has already recovered from previous errors when another error occurs. 
This assumption requires that the probability that an error occurs 
when the decoder has not returned to the synchronization state $n_s$
is very low. 
The lower the error span is and the higher $E_b/N_0$ is, 
the more accurate this approximation is. 
Under this assumption, the quantity $\Delta S$ is independently impacted by multiple errors. 

Let us define
\begin{equation}
G_e(y) = (\underbrace{G \star ... \star G}_{e \text{ times}})(y) = \sum_{i \in \mathbb{Z}} a_{i,e}y^i, 
\end{equation}
where $\star$ denotes the convolution product.
Note that the polynomial $G_1=G$ corresponds to the gain polynomial of Eqn.~\ref{equ:gain_polynomial}. 
Under the previous assumption, the quantity $a_{i,e}$ equals $\P(\Delta S = i | E = e)$.
With Eqn.~\ref{equ:PEe}, the resulting gain polynomial 
for this crossover probability can be expressed as 
\begin{align}
{\tilde G}(y) & = \sum_{e \in \mathbb N} G_e(y) \, \P(E=e), 
\end{align}
where only the quantity $\P(E=e)$ depends on $E_b/N_0$. 
The coefficients ${\tilde {g}_i}$ of $\tilde G$ verify
\begin{align}
{\tilde{g}_i} & = \sum_{e \in \mathbb{N}} a_{i,e} \, \P(E=e) \\
& = \sum_{e \in \mathbb{N}} \P(\Delta S = i | E = e) \P(E=e) \\
& = \P(\Delta S = i).
\label{equ:gi}
\end{align}

 
Let $\eta>0$ be a criterion of negligibility. 
For a given $\eta$, the pseudo-degree $d_{\eta}$ of the polynomial $\tilde{G}$ 
is defined as
\begin{equation}
d_{\eta} \equaldef \min_{\mathbb{N^*}} \, d \, \, {\Big |} {\sum_{i \in {\mathbb Z}-\{-d,...,d\}} \tilde{g}_i < \eta }.
\label{equ:psdeg}
\end{equation}
The pseudo-degree $d_{\eta}$ of a polynomial is the degree beyond which the sum of the coefficients of this polynomial
are below a given threshold $\eta$.

\begin{example}
\label{ex:pseudodegree}
Let us determine the pseudo-degree such that $\eta= 10^{-6}$ 
for the code $\Cset_{5}$, $E_b/N_0 = \text{6\,dB}$, and $L(\S) = 100$. 
The estimates of $\tilde{g}_i$ obtained from Eqn.~\ref{equ:gi} lead to 
\begin{equation}
{\small
\left\{
\setlength{\extrarowheight}{-7pt}
\begin{array}{ll}
\P(\Delta S \leq -4)  & = 0.0000002 \\
\P(\Delta S = -3)     & = 0.0000235 \\
\P(\Delta S = -2)     & = 0.0013201 \\
\P(\Delta S = -1)     & = 0.0493389 \\
\P(\Delta S = 0)      & = 0.9186664 \\
\P(\Delta S = 1)      & = 0.0301524 \\
\P(\Delta S = 2)      & = 0.0004930 \\
\P(\Delta S = 3)      & = 0.0000053 \\
\P(\Delta S \geq 4)   & = 0.0000001.
\end{array}
\right.
}
\end{equation}
Hence, according to the definition of the pseudo-degree in Eqn.~\ref{equ:psdeg},
$d_{\eta} = 3$.  The values of $\P(\Delta S = i)$ obtained by simulation
and averaged over $10^7$ channel realizations are
\begin{equation}
{\small 
\left\{
\setlength{\extrarowheight}{-7pt}
\begin{array}{ll}
\P(\Delta S \leq -4)  & = 0.0000002 \\
\P(\Delta S = -3)     & = 0.0000207 \\
\P(\Delta S = -2)     & = 0.0012587 \\
\P(\Delta S = -1)     & = 0.0500770 \\
\P(\Delta S = 0)      & = 0.9185508\\
\P(\Delta S = 1)      & = 0.0296306\\
\P(\Delta S = 2)      & = 0.0004578 \\
\P(\Delta S = 3)      & = 0.0000041 \\
\P(\Delta S \geq 4)   & = 0.0000001
\end{array}
\right.
}
\end{equation}
and also lead to $d_{\eta} = 3$.
\end{example}\\
The simulated values of $\P(\Delta S = i)$ are close to the estimated ones
for a large set of $E_b/N_0$ values, which validates the approximation. 
The pseudo-degrees for $\eta = 10^{-6}$ of the codes introduced in Table~\ref{tab:codesdef} 
have been computed and are given in Table~\ref{tab:recap_codes}. 

\begin{table*}
\caption{Pseudo-degrees $d_{\eta}$ for $\eta=10^{-6}$, proposed criteria, criteria of~\cite{ZhZ02}, and 
error resilience performance for $E_b/N_0=\textrm{\lowercase{6\,dB}}$, and $L(S)=100$.}
\label{tab:recap_codes}
\begin{center}
\scriptsize
\setlength{\extrarowheight}{2pt}
$$
\begin{array}{|c|c|cc|cr|ccc|}
\hline        
\textrm{Code}  & d_{\eta}  & \P(\Delta S = 0) &   H(\Delta S)   & \textrm{MEPL} \cite{ZhZ02} & \textrm{VEPL} \cite{ZhZ02} 
&  \textrm{NLD}  &  \textrm{BER}    &  \textrm{FER}   \\
\hline
\Cset_{1}    &     3   &  0.9185 &  0.499 & 3.89256 & 34.721 & 0.00877  &  0.00193 &  0.34053 \\
\Cset_{2}    &     4   &  0.9005 &  0.578 & 2.02273 & 2.003  & 0.00632  &  0.00191 &  0.33641 \\
\Cset_{3}    &     4   &  0.8971 &  0.595 & 2.06061 & 2.107  & 0.00626  &  0.00192 &  0.33636 \\ 
\Cset_{4}    &     4   &  0.8913 &  0.608 & 4.07692 & 27.800 & 0.00759  &  0.00177 &  0.31548 \\
\Cset_{5}    &     3   &  0.9187 &  0.497 & 1.71023 & 1.200  & 0.00586  &  0.00194 &  0.34296 \\
\Cset_{6}    &     4   &  0.8996 &  0.578 & 3.54546 & 18.854 & 0.00758  &  0.00182 &  0.32368 \\
\Cset_{7}    &     5   &  0.7088 &  1.287 & 1.55556 & 0.370  & 0.00619  &  0.00154 &  0.21849 \\
\Cset_{8}    &     10  &  0.7006 &  1.553 & 2.34861 & 2.045  & 0.00646  &  0.00134 &  0.19543 \\
\Cset_{9}    &     9   &  0.6703 &  1.632 & 1.95707 & 1.025  & 0.00571  &  0.00123 &  0.16739 \\
\Cset_{10}   &     36  &  0.6401 &  2.267 & 6.18182 & 36.231 & 0.00483  &  0.00074 &  0.10354 \\
\Cset_{11}   &     8   &  0.8797 &  0.655 & 1.85227 & 2.233  & 0.00614  &  0.00183 &  0.32219 \\
\Cset_{12}   &     8   &  0.8882 &  0.620 & 1.71678 & 1.506  & 0.00617  &  0.00187 &  0.32951 \\
\Cset_{13}   &     8   &  0.8860 &  0.634 & 1.79798 & 1.914  & 0.00615  &  0.00182 &  0.32142 \\
\Cset_{14}   &     8   &  0.8957 &  0.599 & 2.03104 & 2.952  & 0.00666  &  0.00186 &  0.32698 \\
\Cset_{15}   &     8   &  0.8941 &  0.610 & 2.20321 & 4.144  & 0.00685  &  0.00189 &  0.33244 \\
\Cset_{16}   &     6   &  0.9044 &  0.564 & 1.98086 & 2.615  & 0.00672  &  0.00193 &  0.33829 \\
\hline
\end{array}
$$
\end{center}
\end{table*}

\subsection{Code selection criteria}
\label{sec:designcriteria}
Let us consider a MAP estimation run on the bit/symbol trellis, with an additional constraint on the length of 
the decoded sequence. This length constraint is used to discard all decoded sequences having a number of symbols which 
differs from the number of transmitted symbols, that is, which does not satisfy the constraint $\Delta S = 0$.
On the bit/symbol trellis, the decoder has two kinds of information to help the estimation: the excess rate of the code and 
the information brougth by the length constraint. The excess rate of a VLC (residual redundancy in the encoded bitstream) is
given by the difference between the mean description length (mdl) of the code and the entropy of the source. The information 
brougth by the length constraint on the bit/symbol trellis is given by the entropy of the p.m.f. of the gain/loss measure
($\Delta S$). For the considered set of codes, the excess rate is equal to $0.0781$ bits of information and is the same for
all codes of Table~\ref{tab:codesdef}.

From the p.m.f. of $\Delta S$ the following two quantities can be computed:
\begin{itemize} 
\item the probability $\P(\Delta S =0)$ to have a strict sense resynchronization 
\item the entropy $H(\Delta S)$. 
\end{itemize}
If the probability $\P(\Delta S =0)$ is small, the number of ``de-synchronized'' sequences which will be discarded will be high, 
then the probability of detecting and correcting errors increases. This results in increasing the likelihood of the correct 
sequence, hence in decreasing the decoding error rate. As explained below, the entropy $H(\Delta S)$ measures the amount of information
brought by the length constraint on the bit/symbol trellis. To design performance criteria for VLCs, we consider codes
having the same mdl so that their performance can be fairly compared. Hence, the values $\P(\Delta S =0)$ and $H(\Delta S)$,
computed from the p.m.f. of the gain/loss measure, are indicators of the performance of a VLC when
soft decoding with length constraint is applied at the decoder side.
Table~\ref{tab:recap_codes} shows the values of these two 
quantities for the codes of Table~\ref{tab:codesdef}, together with the MEPL and the VEPL of~\cite{ZhZ02}. The corresponding 
decoding performance in terms of the normalized Levenshtein distance (NLD)~\cite{Lev67}, BER and frame error rate (FER) obtained 
with the bit/symbol trellis, for $E_b$/$N_0$=~6dB and $L(S)=100$, are also given.  It can be observed that the code $C_{10}$ 
gives the largest MEPL and VEPL. Hence, one could expect this code to lead to the worst decoding performance. However, this 
conclusion is valid only when hard decoding is used. When soft decoding with a length constraint is being used, it can 
be observed that the entropy $H(\Delta S)$ better predicts the decoding performance, the code $C_{10}$ giving the best 
performance in this case in terms of FER, BER and NLD. Similarly, the code $\Cset_{5}$ leads to the worst performance 
in terms of BER and FER.
The same observations can be made for longer sequences (see Table~\ref{tab:L(S)} for $L(S)=500$ and $L(S)=1000$). The MEPL and VEPL criteria are well-suited for hard decoding. However, the two quantities  
$\P(\Delta S =0)$ and $H(\Delta S)$ are better suited in the case of soft decoding with length contraints.

\begin{table}
\caption{Pseudo-degrees, proposed criteria and error resilience performance for $E_b/N_0=\textrm{\lowercase{6\,dB}}$, and
for $L(S)=500$ and $L(S)=1000$.}
\label{tab:L(S)}
\begin{center}
\scriptsize
\setlength{\extrarowheight}{-2pt}
$$
\begin{array}{|c|c|cccc|}
\hline
Code & d_{\eta} & \P(\Delta S = 0) & H(\Delta S) & BER & FER \\
\hline
~ & ~ & ~ & L(S)=500 & ~ & ~ \\
\hline
\Cset_5    & 5   & 0.67565  & 1.39229  & 0.002210  & 0.90012 \\
\Cset_7    & 9   & 0.30597  & 2.49437  & 0.002200  & 0.84090 \\
\Cset_{10} & 40  & 0.13111  & 4.38846  & 0.001687  & 0.64636 \\ 
\hline
~ & ~ & ~ & L(S)=1000 & ~ & ~ \\
\hline
\Cset_5    & 7   & 0.49590  & 1.91479  & 0.002290  & 0.99062 \\
\Cset_7    & 14  & 0.19019  & 3.00963  & 0.002275  & 0.98422 \\
\Cset_{10} & 40  & 0.03215  & 4.97712  & 0.001899  & 0.92838 \\ 
\hline
\end{array}
$$
\end{center}
\end{table}

Simulations have also been performed with a larger source alphabet. The English alphabet together with three Huffman codes 
considered for this source in~\cite{MaR85} and~\cite{SwD95} has been used. This source and the corresponding codes are given
in Table~\ref{tab:eng}. These three codes have the same mean description length ($4.1557$ bits). Table~\ref{tab:eng_perf} gives the 
MEPL and VEPL values, as well as the quantities $\P(\Delta S =0)$ and $H(\Delta S)$, for these codes.  It also gives the FER 
and BER MAP decoding performance of these codes on the bit/symbol trellis. The code $C_{17}$ is the worst code in terms of 
MEPL and VEPL, but the best according to our criteria ($\P(\Delta S =0)$ and $H(\Delta S)$).  This is confirmed by the actual 
FER and BER performance of this code when running the MAP decoder with the length constraint.

\begin{table*}
\caption{Source and codes for the english alphabet used in this paper.}
\label{tab:eng}
\begin{center}
\scriptsize
\setlength{\extrarowheight}{2pt}
$$
\begin{array}{|c|c|c|c|c|}
\hline        
\textrm{ASCII Code}  & \textrm{Symbol probability}  & \Cset_{17} \cite{MaR85}  & \Cset_{18} \cite{MaR85}  & \Cset_{19}\cite{SwD95}  \\
\hline
A & 0.08833733 & 0000         & 0100         & 0100        \\
B & 0.01267680 & 011111       & 111101       & 000101      \\ 
C & 0.02081665 & 11111        & 11100        & 01100       \\
D & 0.04376834 & 00010        & 10110        & 01101       \\
E & 0.14878569 & 001          & 000          & 100         \\
F & 0.02455297 & 11100        & 11010        & 00011       \\
G & 0.01521216 & 011101       & 111011       & 001100      \\
H & 0.05831331 & 1000         & 1000         & 1100        \\
I & 0.05644515 & 1001         & 1001         & 1111        \\
J & 0.00080064 & 111010101    & 111111110    & 001110100   \\
K & 0.00867360 & 1110100      & 1111110      & 0011100     \\
L & 0.04123298 & 00011        & 10111        & 00100       \\
M & 0.02361889 & 11110        & 11011        & 01110       \\
N & 0.06498532 & 0110         & 0111         & 0101        \\
O & 0.07245796 & 0100         & 0101         & 1101        \\
P & 0.02575393 & 10111        & 11001        & 01111       \\
Q & 0.00080064 & 1110101000   & 1111111110   & 0011101010  \\
R & 0.06872164 & 0101         & 0110         & 0000        \\
S & 0.05537763 & 1010         & 1010         & 1110        \\
T & 0.09354149 & 110          & 001          & 101         \\
U & 0.02762209 & 10110        & 11000        & 00101       \\
V & 0.01160928 & 111011       & 111110       & 001111      \\
W & 0.01868161 & 011100       & 111010       & 001101      \\
X & 0.00146784 & 11101011     & 11111110     & 00111011    \\
Y & 0.01521216 & 011110       & 111100       & 000100      \\
Z & 0.00053376 & 1110101001   & 1111111111   & 0011101011  \\
\hline
\end{array}
$$
\end{center}
\end{table*}

\begin{table*}
\caption{Proposed criteria, criteria of~\cite{ZhZ02} and decoding performance of the english alphabet codes on the
bit/symbol trellis for $E_b/N_0=\textrm{\lowercase{6\,dB}}$, and $L(S)=100$.}
\label{tab:eng_perf}
\begin{center}
\scriptsize
\setlength{\extrarowheight}{2pt}
$$
\begin{array}{|c|cc|cr|cc|}
\hline        
\textrm{Code}  & \P(\Delta S = 0) &   H(\Delta S)   & \textrm{MEPL} \cite{ZhZ02} & \textrm{VEPL} \cite{ZhZ02} 
&   \textrm{BER}    &  \textrm{FER}   \\
\hline
\Cset_{17}   &   0.7312 &  1.376 & 5.456 & 5.868  & 0.002082  &  0.53768 \\
\Cset_{18}   &   0.8338 &  0.861 & 3.863 & 3.906  & 0.002094  &  0.56607 \\
\Cset_{19}   &   0.8433 &  0.844 & 1.915 & 1.192  & 0.002105  &  0.56900 \\
\hline
\end{array}
$$
\end{center}
\end{table*}

\section{State aggregation}
\label{sec:statemodel}

The above analysis is used to assess the conditions for optimality of MAP decoding with length constraint on the aggregated 
state model described in~\cite{JMG05}. This model keeps track of the symbol clock values modulo a parameter $T$ instead of 
the symbol clock values as on the classical bit/symbol trellis. The state aggregation leads to a significantly reduced 
decoding complexity, as detailled in Section~\ref{ASM_description}.  
In this section, it is shown that, from $d_{\eta}$ (the pseudo-degree of the polynomial representation 
of $\Delta S$), one can derive the minimal value of $T$ required to have nearly optimum decoding performance (i.e. which closely approaches 
the performance obtained with the bit/symbol trellis). For these values of $T$, we show that the amount of information 
conveyed by the length constraint is not significantly altered by state aggregation.

\subsection{Optimal state model}

\begin{figure}[t]
\begin{center}
    \includegraphics[width=4cm]{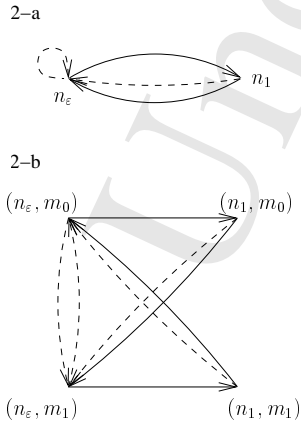}
\end{center}
\caption{Automata of the state models for a) $T=1$, b) $T=2$
corresponding respectively to the bit-level trellis
and the extended trellis with $T=2$ (Code $\Cset_{0}$).}
\label{fig:trellis1}
\end{figure}

\begin{figure*}[t]
\begin{center}
    \includegraphics[width=16cm]{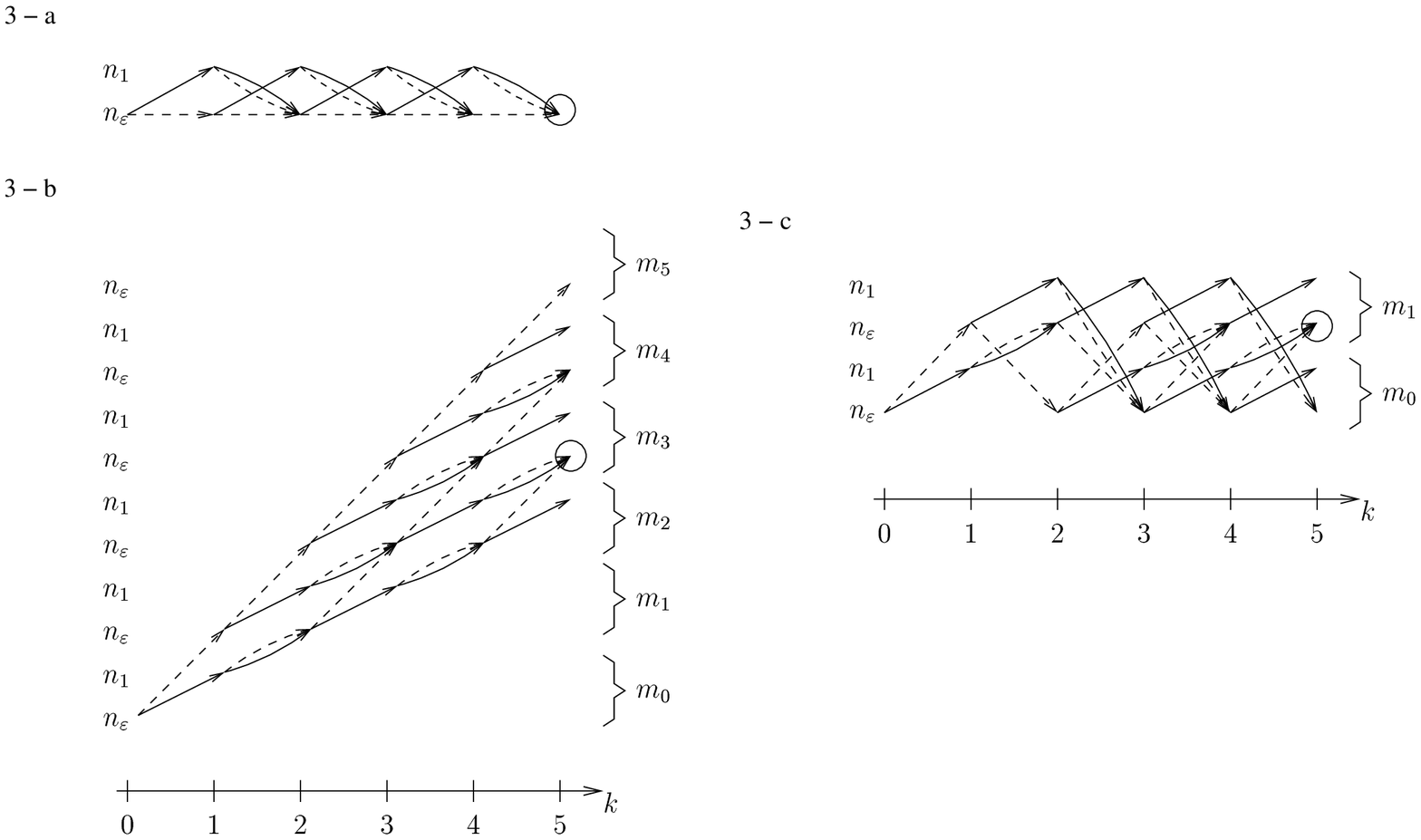}
\end{center}
\caption{Trellises for the code $\Cset_0$:  a) bit-level trellis ($T=1$), b) bit/symbol trellis and c) trellis of parameter $T=2$. 
The termination constraints are also depicted by circles(here $L(\S)$ is assumed to be odd).}
\label{fig:trellis2}
\end{figure*}

The sequence of transmitted bits can be modeled as a hidden markov model with states defined as $X_k$, where $k$ represents 
the bit clock instants, $1 \leq k \leq L(\X)$. Let $N_k$ denote the random variable corresponding to the internal state of 
the VLC (i.e. the internal node of the VLC codetree) at the bit clock instant $k$. For instance, the possible values of $N_k$ 
for the code $\Cset_{0} = \{ 0, 10, 11 \}$ are $n_{\varepsilon}$ and $n_{1}$, where $n_{\varepsilon}$ represents the root node 
of the VLC codetree.  In the bit-level trellis \cite{Bal97}, the decoder state model is defined by the random variable $N_k$ only. 
The internal states of the automaton associated with a given VLC are defined by the internal nodes of the codetree, as depicted in 
Fig.~\ref{fig:trellis1}-a for the code $\Cset_0$. The corresponding decoding trellis is given in Fig.~\ref{fig:trellis2}-a.

Let us assume that the number of transmitted symbols is perfectly known on the decoder side. To use this information as a 
termination constraint in the decoding process, the state model must keep track of the symbol clock (that is of the number 
of decoded symbols).  The optimal state model is defined  by the pair of random variables $(N_k,T_k)$~\cite{gfgr00}~\cite{BaH00e}, 
where $T_k$ denotes the symbol clock instant corresponding to the bit clock instant $k$.  Since the trellis corresponding to this 
model is indexed by both the bit and the symbol instants, it is often called the \emph{bit/symbol} trellis. This trellis is depicted 
in Fig.~\ref{fig:trellis2}-b for the code $\Cset_0$.  The number of states of this model is a quadratic function of the sequence 
length (equivalently the bitstream length).  The resulting computational cost is thus not tractable for typical values of the sequence length $L(\S)$. 

\subsection{Aggregated State model: a brief description}
\label{ASM_description}
The aggregated state model proposed in~\cite{JMG05} is defined by the pair of random variables $(N_k,M_k)$, where $M_k=T_k \mod T$ is the remainder of the 
Euclidean division of $T_k$ by $T$. The corresponding realization of $M_k$ is denoted $m_k$. 
Note that $T=1$ and $T=L(\S)$ amounts to considering respectively the bit-level trellis
and the bit/symbol trellis. The automaton and decoding trellis of parameter $T=2$ corresponding to this state model 
are depicted for the code $\Cset_0$ in Figs.~\ref{fig:trellis1}-b and~\ref{fig:trellis2}-c respectively. The transitions which 
terminate in the state $n_{\varepsilon}$, that is corresponding to the encoding/decoding of a symbol, 
modify $M_k$ as $M_k=M_{k-1} + 1 \mod T$. 
Hence, the transition probabilities on this automaton are given by
\begin{align}
\nonumber
& \P(N_k = n_k , M_k = m_k | N_{k-1} = n_{k-1} , M_{k-1} = m_{k-1} ) = \\
& \left\{
\begin{array}{ll}
\P( N_k=n_k | N_{k-1}=n_{k-1} ) & \text{if } n_k \neq n_{\varepsilon} \text{ and } \\ &  m_k=m_{k-1} \hfill \\
\P( N_k=n_k | N_{k-1}=n_{k-1} ) & \text{if } n_k = n_{\varepsilon} \text{ and }\\ &  m_k = m_{k-1}+1 \text{ mod } T \hfill \\
0  & \text{otherwise} 
\end{array}
\right.
\label{equ:pt}
\end{align} 
where the probabilities $\P(N_k=n_k | N_{k-1}=n_{k-1})$ are deduced from the source statistics. Note that the transition 
probabilities $\P(N_k|N_{k-1})$ are the ones used in the bit-level trellis. 

The proposed state model keeps track of the symbol clock values modulo $T$ during the decoding process. In order to exploit 
this information, the decoder has to know the value $m_{L(\X)}=L(\S) \mod T$. This information can be used as a termination 
constraint, as depicted in Fig.~\ref{fig:trellis2}. If this value is not given by the syntax elements of the source coding 
system, it has to be transmitted. The transmission cost of $m_{L(\X)}$ is greater than or equal to $\log_2(T)$ bits. Note 
that the knowledge of this value has a lower cost than the one of transmitting the exact number of emitted symbols in the 
bit/symbol trellis. In the following, the quantity $m_{L(\X)}$ is assumed to be known by the decoder. The estimation is 
performed using the Viterbi algorithm \cite{Vit67}, hence minimizing the FER. In the sequel, the error resilience will be 
measured according to this criterion. For the estimation, the paths which do not satisfy the appropriate boundary constraints, 
i.e. the paths that do not terminate in states of the form $(n_{\varepsilon},m_{L(\X)})$, are discarded. 
The number of states of the trellis of parameter $T$ satisfies 
\begin{equation}
\nu_T \leq T \times L(\X) \times \Gamma,
\label{equ:nb_states}
\end{equation}
where $\Gamma$ represents the number of internal nodes of the code.
The inequality in Eqn.~\ref{equ:nb_states} results from the fact that some pairs $(n_k,t_k)$ are not reachable according to the
code structure. Such states are mostly located at the first and last bit clock instants of the trellis. However, for
some particular codes, some states are not reachable all along the trellis. For example, for the
set of codewords $\{0,100,101,110,111\}$, the states $(n_{\epsilon},2q),q \in {\mathbb N}$ are not reachable
for any bit clock instants. 
To approximate the complexity on a trellis of parameter $T$, the worst case in terms of
the number of states is considered, i.e. we assume that
\begin{equation}
\nu_T \approx T \times L(\X) \times \Gamma,
\label{equ:nb_states2}
\end{equation}
Hence, as the number of states of the bit-level trellis is equal to
$L(\X) \times \Gamma$, the computational cost $D_{T}$ corresponding to the trellis 
of parameter $T$ can be approximated as 
\begin{equation}
D_T \approx T \times D_{\text{bal}},
\label{equ:approxwithbal}
\end{equation}
where $D_{\text{bal}}$ denotes the computational cost of the bit-level trellis. 
This computational cost is approximatively linear in the sequence length and in $T$.

\subsection{Aggregated state model: analysis}

According to the definition of the pseudo-degree (Eqn.~\ref{equ:psdeg}) of the polynomial $\tilde{G}(y)$, 
the probability that $\Delta S$ belongs to the interval $\{-d_{\eta},\dots,d_{\eta}\}$ is greater than or equal to $1-\eta$. 
This leads to the following property.
\begin{property}
A value of $T$ such that $T = d_{\eta}$, and all the more such that $T > d_{\eta}$, ensures that the Viterbi algorithm run on 
the aggregated trellis selects, with at least probability $1-\eta$, a sequence with the correct number of symbols.
\end{property}

However, this property does not mean that the algorithm will offer similar results as the ones on the bit/symbol trellis. 
To analyze the respective performance of both models, the amount of information conveyed by the termination 
constraint in both cases must be quantified.  These quantities are respectively given by the entropies of the random variables $\Delta S \text{\,mod\,}T$ 
and $\Delta S$. They depend on the sequence length and $E_b/N_0$, which are assumed to be fixed. Here, we show that
by setting the aggregation parameter $T$ to $T=2d_{\eta}+1$, the information brought
by the length constraint on the aggregated trellis ($H(\Delta S \,\text{mod}\, T$)
tends towards the one available on the bit/symbol trellis ($H(\Delta S)$). 

For a trellis of parameter $T$ and following the analysis of Section~\ref{sec:deltaSbsc}, the quantity 
\begin{equation}
\tilde{g}_i^T \equaldef \P(\Delta S \, \text{mod} \, T = i)
\end{equation} 
 can be computed 
from the quantities $\tilde{g_i}$ as
\begin{equation}
 \tilde{g}_i^T = \sum_{j \in \mathbb{Z}} \tilde{g}_{jT+i}. 
\end{equation}
The entropy of the termination constraint on a trellis of parameter $T$ is then given by
\begin{align}
H(\Delta S \, \text{mod \,} T ) & = - \sum_{i \in \{0,\dots,T-1\}} \tilde{g}_i^T \log_2 \tilde{g}_i^T \\
 & \geq H( \Delta S ) + \sum_{i \notin \{0,\dots,T-1\}} \tilde{g}_i \log_2  \tilde{g}_i. 
\label{equ:entropy1}
\end{align}
When $T=2 d_{\eta}+1$, (\ref{equ:entropy1}) can be re-written as
\begin{equation}
H(\Delta S \, \text{mod} \, 2 d_{\eta}+1 ) 
  \geq H( \Delta S ) + \sum_{i \notin \{-d_{\eta},\dots,d_{\eta}\}} \tilde{g}_i \log_2  \tilde{g}_i. 
\end{equation} 

Let us now assume that $\eta < \frac{1}{e}$.  Then the function $x \mapsto x \log(x)$ decreases on the interval $[0,\eta]$ 
and since $\forall i \notin \{-d_{\eta},\dots,d_{\eta}\}$, $\tilde{g}_i \leq \eta$, we have 
\begin{equation}
\sum_{i \notin \{-d_{\eta},\dots,d_{\eta}\}} \tilde{g}_i \log_2  \tilde{g}_i  
  \geq |\{ i \notin \{-d_{\eta},\dots,d_{\eta}\}], \tilde{g}_i > 0 \}| \, \eta \log_2 \eta, 
\end{equation}
where the cardinal $|\{ i, \tilde{g}_i > 0 \}|$ of the set of possible non-zero values of $\tilde{g}_i$ is bounded by the 
bitstream length $L(\X)$. Together with $L(\X) \geq T$, this leads to 
\begin{equation}
|\{ i \notin \{-d_{\eta},\dots,d_{\eta}\}, \tilde{g}_i > 0 \}| \leq L(\X) - 2 \, d_{\eta} - 1. 
\end{equation} 

Hence, for a given $\eta$, we have the following lower and upper bounds:
\begin{equation}
\label{equ:H_ineg1}
H(\Delta S) + (L(\X)-2 \, d_{\eta} - 1) \, \eta \log_2 \eta \leq H(\Delta S \, \text{mod} \, 2\,d_{\eta}+1) \leq H(\Delta S).  
\end{equation}
These bounds mean that for $\eta$ small enough, hence for $T=2\,d_{\eta}+1$ sufficiently high, the quantity of information 
brought by the length constraint on the aggregated trellis of parameter $T$ tends toward the one available on the bit/symbol trellis. 

\begin{example}
Let us consider the same parameters as in Example~\ref{ex:pseudodegree} (i.e. code $\Cset_{5}$, $d_{\eta} = 3$). From Eqn.~\ref{equ:H_ineg1}  we deduce that 
\begin{align}
\nonumber
H(\Delta S) - H(\Delta S \, \text{mod} \, 2\,d_{\eta}+1) & \leq -(100 \, l_M-5) \eta \log_2(\eta)  \\
& \leq 0.04900 \text{ bits}
\end{align}
\end{example}

\begin{figure}[t]
\begin{center}
     \includegraphics[width=9cm]{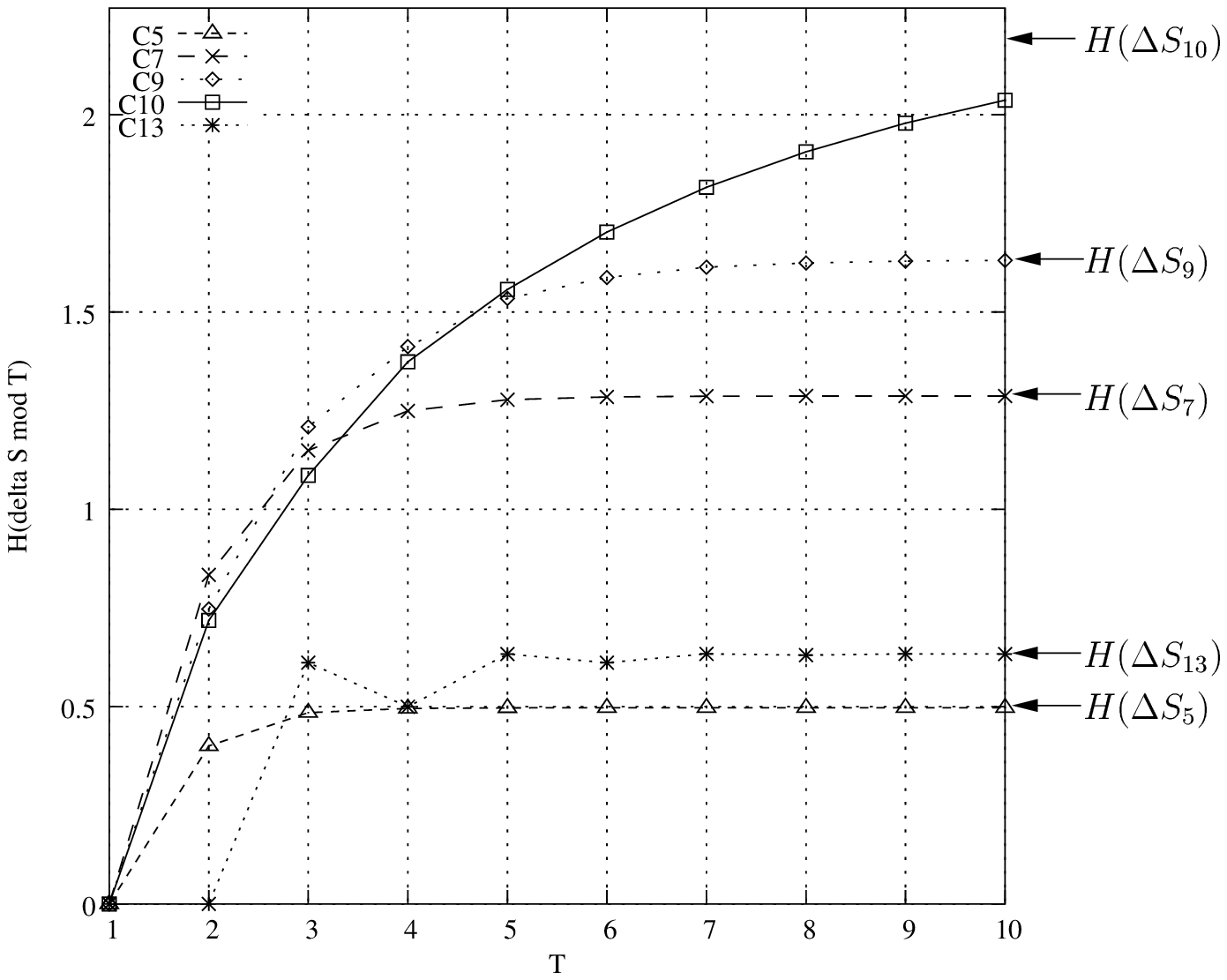}
\end{center}
\caption{Entropy of $\Delta S \, \text{mod} \, T$ versus $T$ for the codes $\Cset_{5}, \Cset_{7} , \Cset_{9}, \Cset_{10}$ 
and $\Cset_{13}$.}
\label{fig:HdeltaSmod}
\end{figure}
The convergence of $H(\Delta S \, \text{mod} \, T)$ is depicted in Fig.~\ref{fig:HdeltaSmod} for codes of Table~\ref{tab:codesdef}.
In this figure, the arrows represent the values of $H(\Delta S)$ for the considered codes. Note that for $\Cset_{10}$,
$H(\Delta S \, \text{mod} \, T)$ has not converged towards $H(\Delta S)$ yet for $T=10$. For the other codes, the
limit is reached for $T \leq 10$. 
According to Section~\ref{sec:designcriteria}, the best codes are those with the highest values of $H(\Delta S)$. 
Such codes require a higher value of $T$ to approach the value $H(\Delta S)$ of the entropy of the termination constraint on the 
bit/symbol trellis, since the pseudo-degree of these codes is higher.  Nevertheless, for the considered set of codes, the 
values of $T$ leading to the same performance as on the bit/symbol trellis are always lower than $L(\X)$. 
Note that for the code $\Cset_{13}$, $H(\Delta S \, \text{mod} \, 2) = 0 = H(\Delta S \, \text{mod} \, 1)$.
This means that the decoding performance of code $\Cset_{13}$ on a trellis of parameter $T=2$ is the
same as the one on the bit/level trellis ($T=1$).


The previous analysis has been validated by simulation, for sequences of $L(\S)=100$ symbols. For each parameter 
set (VLC, $E_b/N_0$ and $T$),  the FER is measured over $10^5$ channel realizations. 
The performance at different values of the parameter $T$ and for the codes $\Cset_{5}, \Cset_{7} , \Cset_{10}$ 
and $\Cset_{13}$ is given in Table~\ref{tab:sqer}. In this table, the best decoding performance for each code, 
at different values of $E_b/N_0$ is written in \textit{italics}.  These values correspond to the performance 
obtained on the bit/symbol trellis. 
Note that these values are obtained for a value of $T$ which is considerably lower than $L(\S)$.
As predicted, the trellis of parameter $T=2$ does not bring any improvement in terms of error resilience for the code
$\Cset_{13}$ compared to the bit-level trellis.
These results validate the criteria described in Section~\ref{sec:designcriteria} to select good
codes in terms of error resilience. Indeed, according to these criteria and
the simulation results, the best code among the ones proposed in Table~\ref{tab:codesdef}
is the code $\Cset_{10}$ and the worst is the code $\Cset_{5}$.



\begin{table}[t]
\caption{FER for soft decoding (Viterbi)
with different values of the aggregation parameter $T$.}
\label{tab:sqer}
\begin{center}
\scriptsize
$$
\begin{array}{|c|ccccc|}
\hline
{E_b}/{N_0} &    3    &     4    &      5     &      6     &      7 \\     
\hline            
~ & ~ & ~ & \textrm{Code } \Cset_{5} & ~ & ~ \\
\hline
T=1   & 0.99120   & 0.92330  & 0.70464  & 0.38774  & 0.14558 \\
T=2   & 0.98805   & 0.90368  & 0.66193  & 0.34633  & 0.12452 \\
T=3   & 0.98698   & 0.89901  & 0.65527  & 0.34313  & 0.12388 \\ 
T=4   & 0.98665   & 0.89795  & 0.65457  & 0.34298  & \textit{0.12386} \\ 
T=5   & 0.98652   & 0.89782  & 0.65449  & \textit{0.34296}  &  \\ 
T=10  & \textit{0.98651}   & \textit{0.89780}  & \textit{0.65448}  &   &  \\
\textrm{bit/symb.}(T=100)         & 0.98651   & 0.89780  & 0.65448  & 0.34296  & 0.12386 \\
\hline
~ & ~ & ~ & \textrm{Code } \Cset_{7} & ~ & ~ \\
\hline
T=1           & 0.99182   & 0.92604  & 0.71405  & 0.39372  & 0.14885 \\ 
T=2           & 0.98634   & 0.88506  & 0.59864  & 0.25742  & 0.06997 \\
T=3           & 0.98247   & 0.86379  & 0.55406  & 0.22571  & 0.06152 \\
T=4           & 0.98005   & 0.85387  & 0.53964  & 0.21947  & 0.06059 \\
T=5           & 0.97893   & 0.84960  & 0.53581  & 0.21866  & \textit{0.06057} \\
T=10          & 0.97773   & \textit{0.84731}  & \textit{0.53468}  & \textit{0.21849}  &  \\
T=20          & \textit{0.97772}   &   &  &  &  \\
\textrm{bit/symb.}(T=100)   & 0.97772   & 0.84731  & 0.53468  & 0.21849  & 0.06057 \\
\hline
~ & ~ & ~ & \textrm{Code } \Cset_{10} & ~ & ~ \\
\hline
T=1      & 0.97993   & 0.87316  & 0.61783  & 0.31353  & 0.11390 \\ 
T=2      & 0.96917   & 0.82122  & 0.51758  & 0.22232  & 0.06832 \\
T=3      & 0.96092   & 0.78516  & 0.46126  & 0.18023  & 0.05207 \\
T=4      & 0.95331   & 0.75512  & 0.41127  & 0.14437  & 0.03718 \\
T=5      & 0.94755   & 0.73502  & 0.38403  & 0.12851  & 0.03226 \\
T=10     & 0.93238   & 0.68744  & 0.33174  & 0.10496  & 0.02631 \\
T=20     & 0.92801   & 0.67825  & 0.32560  & \textit{0.10354}  & \textit{0.02610} \\
T=30     & \textit{0.92791}   &\textit{0.67811}  & \textit{0.32558}  &  &  \\ 
\textrm{bit/symb.}(T=100)     & 0.92791   & 0.67811  & 0.32558  & 0.10354  & 0.02610 \\
\hline
~ & ~ & ~ & \textrm{Code } \Cset_{13} & ~ & ~ \\
\hline
T=1    & 0.98973   & 0.91752  & 0.69351  & 0.38031  & 0.14431 \\ 
T=2    & 0.98973   & 0.91752  & 0.69351  & 0.38031  & 0.14431 \\
T=3    & 0.98369   & 0.88547  & 0.62816  & 0.32182  & 0.11644 \\
T=4    & 0.98552   & 0.89259  & 0.63858  & 0.32711  & 0.11762 \\
T=5    & 0.98286   & 0.88356  & 0.62642  & \textit{0.32142}  & \textit{0.11638} \\
T=10   & 0.98286   & 0.88356  & 0.62642  &  & \\
T=20   & \textit{0.98277}   & \textit{0.88348}  & \textit{0.62638}  &  & \\
\textrm{bit/symb.}(T=100)  & 0.98277   & 0.88348  & 0.62638  & 0.32142  & 0.11638 \\
\hline
\end{array}
$$
\end{center}
\end{table}

\section{Combined trellis Decoding}
\label{sec:mtd}
\subsection{Motivation}
In this section, we propose an approach allowing further reduction of the decoding complexity without inducing any suboptimality in
terms of decoding performance. 
The optimality of this approach is proved for the FER criterion.
This approach is motivated by the following equivalence
\begin{align}
\nonumber
L(\S) & \mod (T_1 \times T_2) = m  \\
\Leftrightarrow 
& \left\{
\begin{array}{l}
L(\S) \mod T_1 = m \mod T_1 \\
L(\S) \mod T_2 = m \mod T_2, \\
\end{array}
\right.
\label{equ:modproj}
\end{align} 
satisfied if $T_1$ and $T_2$ are relatively prime.
Note that, if $T_1$ and $T_2$ are not relatively prime, the converse 
is not satisfied.

\begin{property}
Let us assume that $T_1$ and $T_2$ are relatively prime and that 
$T_3 \equaldef T_1 \times T_2$.
Let us denote by $\hat{\S}_1$, $\hat{\S}_2$ and $\hat{\S}_3$ the estimates 
of $\S$ provided by the Viterbi algorithm run 
on the trellises of parameters $T_1$, $T_2$ and $T_3$ respectively. 
Then, we have 
\begin{equation}
\hat{\S}_1 = \hat{\S}_2 \Rightarrow \hat{\S}_3 =\hat{\S}_1 = \hat{\S}_2. 
\end{equation}
\label{equ:property_seqvalid}
\end{property}
\begin{proof}
Let us first emphasize that the probability of a sequence, computed by
the Viterbi algorithm on a trellis of parameter $T$ does not depend on $T$. 
Let us assume that if two sequences have the same probability, then 
a subsidiary rule is applied to select one sequence amongst the two. 
For instance, the lexicographical order can be chosen as a comparison rule. 
Such a rule ensures that the Viterbi algorithm behavior is deterministic. 
Let 
\begin{equation}
{\mathcal S}_T \equaldef \{ s' / L(s') \text{\,mod\,} T = L(s) \text{\,mod\,} T \}
\end{equation}
be the set of sequences satisfying the termination constraint for the trellis 
of parameter $T$. 
From Eqn.~\ref{equ:modproj}, we deduce that if $T_3=T_1 \times T_2$ with 
$T_1$ and $T_2$ relatively prime, then 
\begin{equation}
{\mathcal S}_{T_3} = {\mathcal S}_{T_1} \cap {\mathcal S}_{T_2}, 
\label{equ:S3S1S2}
\end{equation}
hence, 
\begin{equation}
{\mathcal S}_{T_3} \subseteq {\mathcal S}_{T_1}.
\end{equation}

Moreover, since we have assumed that $\hat{\mathbf S}_1 = \hat{\mathbf S}_2$, we get
\begin{equation}
\hat{\mathbf S}_1 \in {\mathcal S}_{T_3}. 
\end{equation}

The estimate $\hat{\S}_{1}$ provided by the Viterbi algorithm applied on the trellis 
of parameter $T_1$ is then such that 
\begin{align}
\hat{\mathbf S}_{1} & = \text{arg} \max_{s' \in {\mathcal S}_{T_1}} \P( s' | {\mathbf X}) \\
 & = \text{arg} \max_{s' \in {\mathcal S}_{T_3}} \P( s' | {\mathbf X}) \\ 
 & = \hat{\mathbf S}_{3}, 
\end{align} 
where the subsidiary rule may be used in the selection of the maximum. 
This concludes the proof.  
\end{proof}

This property means that if a sequence is selected by the trellises of parameters 
$T_1$ and $T_2$, then this sequence is also selected by the trellis of parameter $T_3$.

\subsection{The decoding algorithm}
\label{sec:decoding_algorithm}

The purpose of the algorithm described in this section is to exploit Property~\ref{equ:property_seqvalid}. 
The corresponding approach is referred to as \emph{combined trellis decoding}. The rationale behind this approach 
is to use two trellises of parameters $T_1$ and $T_2$ instead of the trellis of parameter $T=T_1 \times T_2$ in 
order to reduce the overall decoding complexity. We will also assume that the greatest common divisor (gcd) of $T_1$ and $T_2$ is $1$, 
i.e. that $T_1$ and $T_2$ are relatively prime.  The decoding of a sequence proceeds as follows: 
\begin{enumerate}
\item The Viterbi algorithm is applied to both trellises $T_1$ and $T_2$.
They respectively provide the estimated sequences $\hat{\S}_1$ and $\hat{\S}_2$. 
\item If $\hat{\S}_1 = \hat{\S}_2$, 
the decoded sequence is used as the estimate of the emitted sequence.
\item Else, the Viterbi algorithm is applied to the trellis
of parameter $T_1 \times T_2$. 
\end{enumerate}
According to Property~\ref{equ:property_seqvalid},  if the same sequence is selected by both trellises $T_1$ and $T_2$,
this sequence is also selected by the trellis of parameter $T_1 \times T_2$. 
Hence, the performance of the above 3-step decoding algorithm of parameters $T_1$ and $T_2$   
is equivalent to the one obtained with a Viterbi decoder operating on the trellis of parameter $T_1 \times T_2$. 

\subsection{Expected computational cost of the proposed algorithm}

\begin{figure}[t]
\begin{center}
   \hspace{-1cm}   \includegraphics[width=9.5cm]{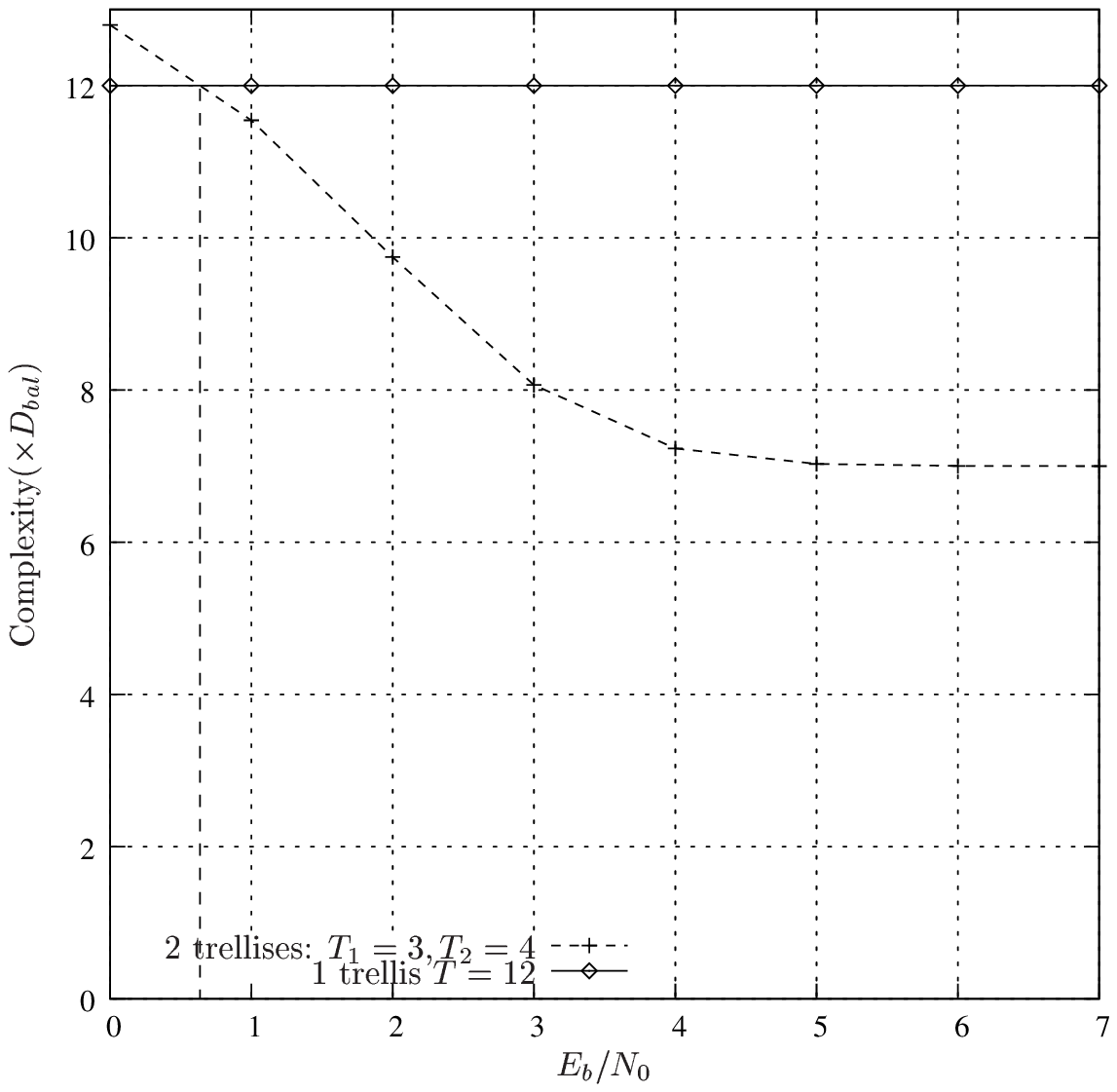} 
\end{center}
\caption{Computational cost of the combined trellis decoding approach 
versus $E_b/N_0$ against the computational cost of a single trellis decoding approach
for parameters $(T_1=3,T_2=4,T_3=12)$.
The corresponding cut-off value of $E_b/N_0$ is also depicted
and is obtained for $\rho^*=0.417$.}
\label{fig:rho}
\end{figure}

First, let us recall that if $T=1$, the resulting trellis is equivalent to the bit-level trellis.
If $T$ is greater than or equal to $L(\S)-\frac{L(\X)}{l_M}+1$ (hence greater than or equal to $L(\S)$), 
the trellis is equivalent to the bit/symbol trellis.  The intermediate values of $T$ amount to considering trellises 
whose complexity is lower than the one of the bit/symbol trellis (see Section \ref{ASM_description}). 
The expectation $D_{\text{mtd}}(T_1,T_2)$ of the computational cost of the proposed decoding scheme 
is then given by
\begin{equation}
D_{\text{mtd}}(T_1,T_2) = T_1 D_{\text{bal}} + T_2 D_{\text{bal}} + \rho T_1 T_2 D_{\text{bal}} 
\end{equation}
where $\rho = \P(\hat{\S}_1 \neq \hat{\S}_2)$. 
In the following, $D_{\text{mtd}}(T_1,T_2)$ will be denoted $D_{\text{mtd}}$. 
The proposed method is worthwhile in terms of computational cost if 
$D_{\text{mtd}} < T_1 \times T_2 \times D_{\text{bal}}$, 
i.e. if
\begin{equation}
\rho < \rho^* = 1 - \frac{T_1 + T_2}{T_1 \times T_2}. 
\end{equation}
Therefore, the benefit of the proposed algorithm depends on the probability~$\rho$ that the two estimators return the same 
sequence estimate. The probability $\rho$ decreases when the channel noise and/or the sequence length increases.  
Fig.~\ref{fig:rho} illustrates the  complexity reduction brought by the combined trellis decoding 
algorithm for the same decoding performance. For the considered settings, a lower computational 
cost is obtained with this approach as long as ${E_b}/{N_0}$ 
is greater than 0.65\,dB.

\subsection{Constrained optimization of trellis parameters $T_1$ and $T_2$}

Let $T_c$ be a targeted decoding performance. According to the combined trellis
decoding scheme described above, this level of performance can be reached
using two trellises of parameters $T_1$ and $T_2$ such that $T_1 \times T_2 = T_c$,
$T_1$ and $T_2$ being relatively prime.
Without loss of generality, let us assume that $T_2=T_1+\Delta T$, and
$T_c = T_1 \times (T_1 + \Delta T)$.
Note that parsing the set ${\mathbb N}^* \times{\mathbb N}^* $ 
with the pairs $(T_1,\Delta T)$ ensures to parse the set of attainable constraints.
The probability $\rho$ is a function of $T_1$ and $T_2$, hence a function
of $T_1$ and $\Delta T$.
The computational cost $D_{mtd}$ of the combined trellis decoding algorithm of
parameters $T_1$ and $T_1+\Delta T$ is given by
\begin{align}
D_{\text{mtd}}(T_1,\Delta T) & = \rho(T_1,\Delta T) D_{T_c} + D_{\text{bal}} (2 \, T_1 + \Delta T )
\label{equ:Dmtdequal} 
\end{align} 

The quantity $\rho(T_1,\Delta T)$ represents the probability that the trellises of parameter $T_1$ and $T_1+\Delta T$ do not
provide the same estimate. This quantity can hence be assumed to increase with $\Delta T$. This assumption may not be satisfied 
for codes having specific synchronisation recovery properties. For example, according to section~III, even values of $T$ are not 
appropriate for the code $\Cset_{13}$. Indeed, for this code, a trellis of parameter $T=2q-1,q \in {\mathbb N}$ provides better decoding performance than 
a trellis of parameter $T=2q,q \in {\mathbb N}$. The previous assumption is not always satisfied for this specific code. 
Under the assumption that $\rho(T_1,\Delta T)$ increases with $\Delta T$, we deduce the following property from Eqn.~\ref{equ:Dmtdequal}.

\begin{property}
Let $T_c \in {\mathbb N}^*$ and ${\mathcal R}_p \subseteq N^* \times N^*$ 
be the subset of positive integers which are relatively prime. Then 
\begin{equation}
\arg \min_{T_1,T_2 \in {\mathcal R}_p  \ /\ 
T_1 T_2 = T_c 
} D_{\text{mtd}} 
= \arg \min_{T_1,T_2 \in {\mathcal R}_p \ /\ T_1 T_2 = T_c} |T_2 - T_1|. 
\label{equ:mtdbestdelta}
\end{equation}
\end{property}

According to that property, the set of pairs $(T_1,T_2)$ such that $T_2=T_1+1$ is optimum.

\section{Conclusion}

This paper makes the link between re-synchronisation properties of VLCs and 
length-constrained MAP estimation techniques of these codes.  This analysis is also used to assess conditions for 
optimality of state aggregation on the bit/symbol trellis widely used for soft decoding of VLC encoded sources. 
Nearly optimal decoding performance can be achieved with a reduced decoding complexity with respect to the classical bit/symbol 
trellis.  A combined trellis decoding algorithm, further reducing the decoding complexity without inducing suboptimality, 
is then proposed.  The aggregated trellises can easily be coupled with a convolutional code or a turbo-code 
in an iterative structure, as done in~\cite{gfgr00}, and~\cite{BaH00c}.

\bibliographystyle{IEEEtran}
\bibliography{abrv,JG,biblio,softVLC,JSC}

\end{document}